\documentclass[twocolumn,floats,floatfix,prl,superscriptaddress,showpacs,aps]{revtex4-1}
\usepackage{graphicx}%
\usepackage{dcolumn}%
\usepackage{bm}%
\usepackage{chemist}

\begin{document}

 \preprint{APS/123-QED}

\title{Effects of electron-phonon interactions on the electron tunneling spectrum of PbS quantum dots}

\author{H. Wang}
\affiliation{ESPCI-ParisTech, PSL Research University, UPMC Univ. Paris 06,  10 rue Vauquelin, LPEM, CNRS, F-75231 Paris Cedex 5, France}
\author{E. Lhuillier}
\affiliation{ESPCI-ParisTech, PSL Research University, UPMC Univ. Paris 06,  10 rue Vauquelin, LPEM, CNRS, F-75231 Paris Cedex 5, France}
\author{Q. Yu}
\affiliation{ESPCI-ParisTech, PSL Research University, UPMC Univ. Paris 06,  10 rue Vauquelin, LPEM, CNRS, F-75231 Paris Cedex 5, France}
\author{A. Mottaghizadeh}
\affiliation{ESPCI-ParisTech, PSL Research University, UPMC Univ. Paris 06,  10 rue Vauquelin, LPEM, CNRS, F-75231 Paris Cedex 5, France}
\author{C. Ulysse}
\affiliation{Laboratoire de Photonique et de Nanostructures, CNRS, 91460 Marcoussis, France}
\author{A. Zimmers}
\affiliation{ESPCI-ParisTech, PSL Research University, UPMC Univ. Paris 06,  10 rue Vauquelin, LPEM, CNRS, F-75231 Paris Cedex 5, France}
\author{A. Descamps-Mandine}
\affiliation{ESPCI-ParisTech, PSL Research University, UPMC Univ. Paris 06,  10 rue Vauquelin, LPEM, CNRS, F-75231 Paris Cedex 5, France}
\author{B. Dubertret}
\affiliation{ESPCI-ParisTech, PSL Research University, UPMC Univ. Paris 06,  10 rue Vauquelin, LPEM, CNRS, F-75231 Paris Cedex 5, France}
\author{H. Aubin}
\email{Herve.Aubin@espci.fr} 
\affiliation{ESPCI-ParisTech, PSL Research University, UPMC Univ. Paris 06,  10 rue Vauquelin, LPEM, CNRS, F-75231 Paris Cedex 5, France}

\date{\today}

\begin{abstract}
We present a tunnel spectroscopy study of single PbS Quantum Dots (QDs) as function of temperature and gate voltage. Three distinct signatures of strong electron-phonon coupling are observed in the Electron Tunneling Spectrum (ETS) of these QDs. In the shell-filling regime, the $8\times$ degeneracy of the electronic levels is lifted by the Coulomb interactions and allows the observation of phonon sub-bands that result from the emission of optical phonons. At low bias, a gap is observed in the ETS that cannot be closed with the gate voltage, which is a distinguishing feature of the Franck-Condon (FC) blockade. From the data, a Huang-Rhys factor in the range $S\sim 1.7 - 2.5$ is obtained. Finally, in the shell tunneling regime, the optical phonons appear in the inelastic ETS $d^2I/dV^2$.  
\end{abstract}


\maketitle

Semiconducting nanocrystals are characterized by discrete electronic levels with size-tunable energies\cite{Alivisatos1996}, giving these QDs unique electronic properties\cite{Delerue2004,Talapin2005,Urban2007}.

While optical spectroscopy is usually used to characterize the properties of QDs, ETS is a more relevant characterization when the goal is to incorporate the QDs into electron conducting devices such as field-effect transistors\cite{Talapin2005} or light emitting diodes\cite{Chen2013}. Indeed, the coupling of a QD to electrodes or neighboring QDs, in presence of Coulomb and electron-phonon interactions, strongly alters their electronic spectrum and, consequently, their electronic transmission coefficient.

In this work, we have studied the ETS of PbS QDs. They are characterized by strong quantum confinement and a size-tunable band gap on a wide energy range, which is of interest for solar cells \cite{Wise2000,Konstantatos2006,Klimov2007,Pandey2008} and infra-red detectors\cite{Lhuillier2014}. 





After synthesis of the PbS QDs, as described in Ref.\cite{Koh2011,Supp} and shown on the TEM picture Fig.~1a., the organic ligands at their surface are replaced by short inorganic ligands, S$^{2-}$\cite{Nag2011,Lhuillier2014}, to reduce the thickness of the insulating tunnel barrier between the QD and the electrodes.

\begin{figure}[ht!]
	\begin{center}
		\includegraphics[width=7cm]{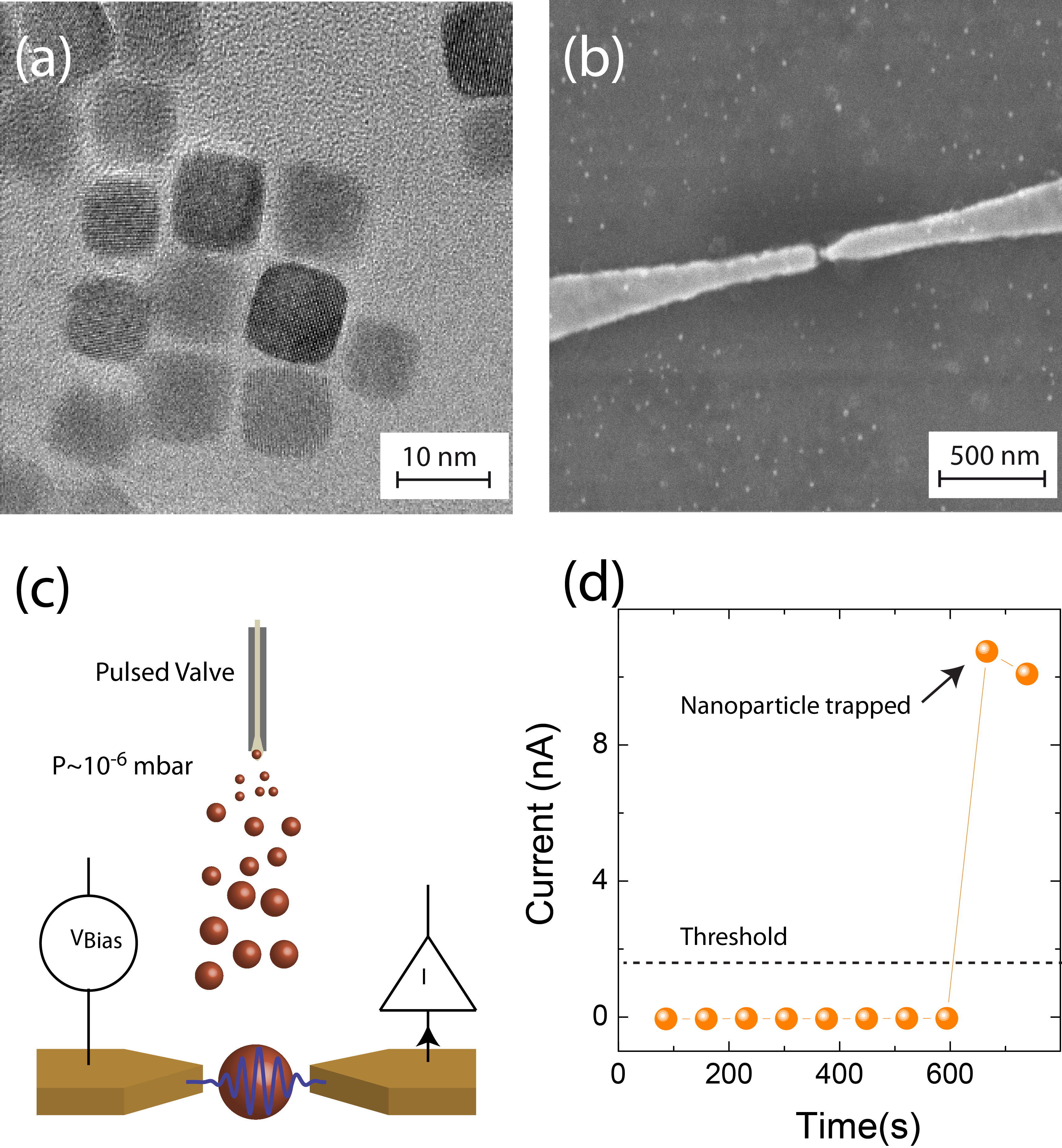}
		\caption{\label{Fig1} a) TEM image of PbS QDs. b) SEM image of $\sim 10~nm$ spaced electrodes in which a QD has been deposited. c) QDs are projected onto the chip-circuit in high vacuum using a fast pulsed valve. d) After each projection, the tunnel current is measured ($V_{Drain}=0.1$~V, $V_{Gate}=0$~V, T=300~K). When it exceeds the threshold, the projection stops.}
	\end{center}
\end{figure}

To measure the ETS as function of temperature and carrier filling, we employed on-chip tunneling spectroscopy where the nanoparticle is trapped within a nanogap, i.e. two electrodes separated by a distance of about $10$~nm, deposited on a p-doped silicon substrate used as a back-gate covered by a silicon oxide layer $300$~nm thick. While Scanning Tunneling Microscopy (STM) has already been employed to study the ETS of several colloidal QDs systems\cite{Millo2000,Bakkers2001,Millo2001,Banin1999,Liljeroth2005,Jdira2008a,Sun2009,Diaconescu2013,Wang2014a}, on-chip tunneling spectroscopy has been only employed a few times\cite{Berkeley1997,Kuemmeth2008,Yu2013a}. This method presents several advantages though. The junctions are highly stable at low temperature, which allows high resolution measurements of the elastic and inelastic ETS. A back gate can be implemented, which allows changing the carrier filling of the QD.

To trap the QDs within the nanogap, we developed a new method\cite{Yu2013b,Yu2013a} where the chip is maintained in high vacuum, $10^{-6}$~mbar, and the QDs are projected through a fast pulsed valve, Fig.~1c. After each projection, the tunnel current is measured to check for the presence of a QD. The projection is repeated hundreds of times until a QD is detected. This generates a \emph{projection curve}, Fig.~1d, where the tunnel current is zero until a QD gets trapped within the nanogap which leads to a sharp increase of the tunnel current. This method has significant advantages. First, because the sample is fabricated in high vacuum, the tunnel current can be measured during the projection of the nanoparticles. Second, the method allows hundreds of trials, i.e. projection-measure, in a few hours, which increase significantly the probability of fabricating single nanoparticle devices. 10 chip circuits have been fabricated and measured from $T=300$~K to $T=5$~K. The projection setup, as well as the cryofree cryostat employed for measurements, are implemented in a glove box under argon. The ETS $dI/dV$ and inelastic ETS $d^2I/dV^2$ are measured with a lock-in. The data for three samples, A, B and C, are shown. 


\begin{figure}[ht!]
	\begin{center}
		\includegraphics[width=8cm]{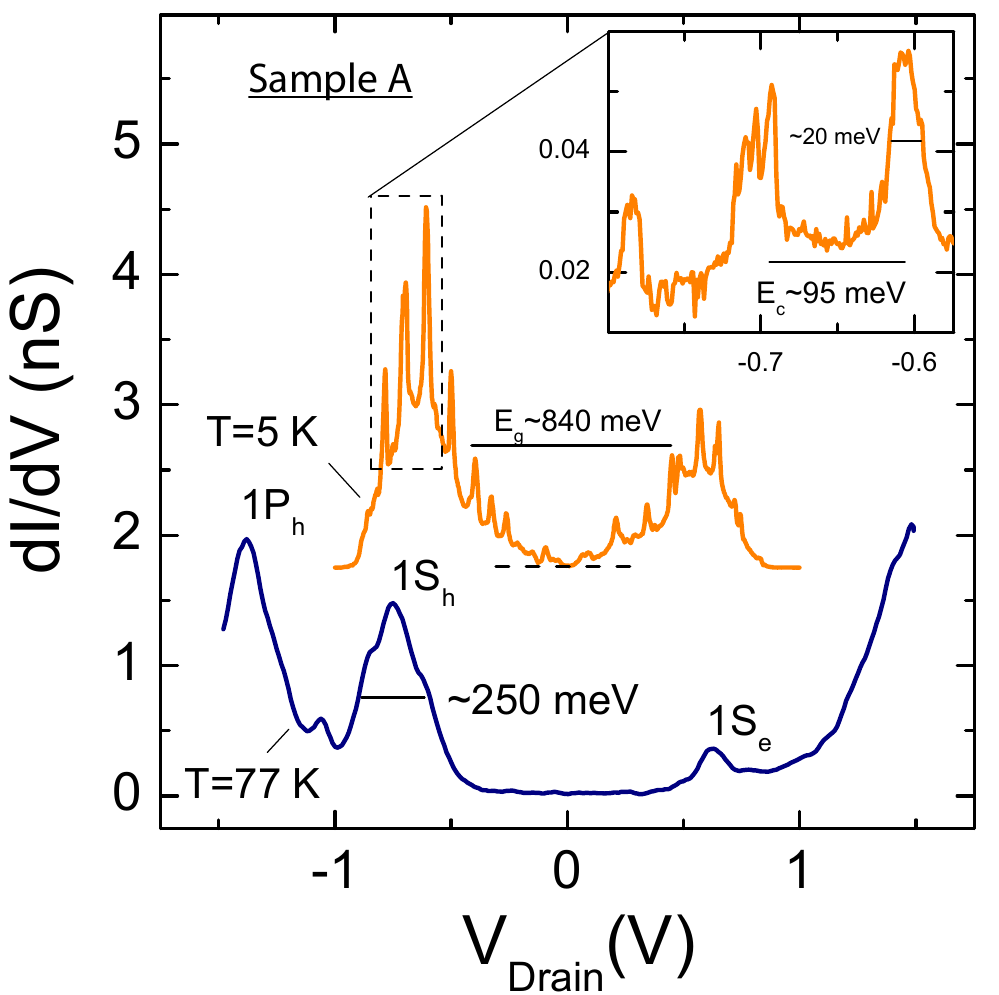}
		\caption{\label{Fig2}
		 $dI/dV$ for sample A. The curve at $T=77$~K shows the main excited levels  1S$_h$, 1P$_h$ and  1S$_e$. The curve at $T=5$~K shows that the degeneracy of the excited levels has been lifted by the Coulomb interactions and gives rise to Coulomb peaks. This last curve has been shifted up for clarity, where the dash line indicates zero level. The inset is a zoom on the Coulomb peaks showing that their width, $\sim 20$~meV, is larger than thermal smearing $\sim 0. 45$~meV. For these measurements, $V_{Gate}=0$~V.}
	\end{center}
\end{figure}

Figure 2 shows the $dI/dV$ curves measured on sample A at two different temperatures.
At the highest temperature, $T=77$~K, the curve shows conductance peaks corresponding to the excited hole levels 1S$_h$, 1P$_h$ and electron level 1S$_e$ of the QD.

At the lower temperature, $T=5$~K, the ETS is modulated by sharp conductance peaks which are characteristics Coulomb blockade peaks in the shell filling regime\cite{Banin2003,Jdira2008b}. In this regime, the tunneling rate $\Gamma_{in}$ for electrons entering the QD is larger than the tunneling rate $\Gamma_{out}$ for electrons escaping the QD. From the voltage separation between two peaks, we obtain the value $E_c\sim 95$~meV for the Coulomb energy.


This experimental value is consistent with the calculated Coulomb energy $E_c=e^2/C_{self}$ where $C_{self}=r/(1/\kappa_m+0.79/\kappa_{PbS})$ is the self-capacitance of the QD, using for the diameter $2\times r\sim 8.5$~nm, $\kappa_m=4\pi \varepsilon_m\varepsilon_0$ with $\varepsilon_m=1.8$, which is the average dielectric coefficient of the media surrounding the QD, and $\kappa_{PbS}=4\pi \varepsilon_{PbS}\varepsilon_0$ where $\varepsilon_{PbS}=170$ is the static dielectric coefficient of PbS. This analysis ignores a possibly small contribution of the electrodes to the Coulomb energy.

From these parameters, we  also obtain the polarisation energy\cite{Grinbom2010,Niquet2002,Supp}, $\Sigma\sim 95$~meV. As the excitation gap $E_{g0}$ is related to the tunneling gap  $E_g$ through the relation $E_g=E_{g0}+2 \Sigma$, one find the experimental value $E_{g0} \sim 640$~meV at $T=5$~K. This value is consistent with the excitation gap expected from \textbf{k.p} four bands envelope function formalism\cite{Kang1997}. 

\begin{figure}[ht!]
	\begin{center}
		\includegraphics[width=8cm]{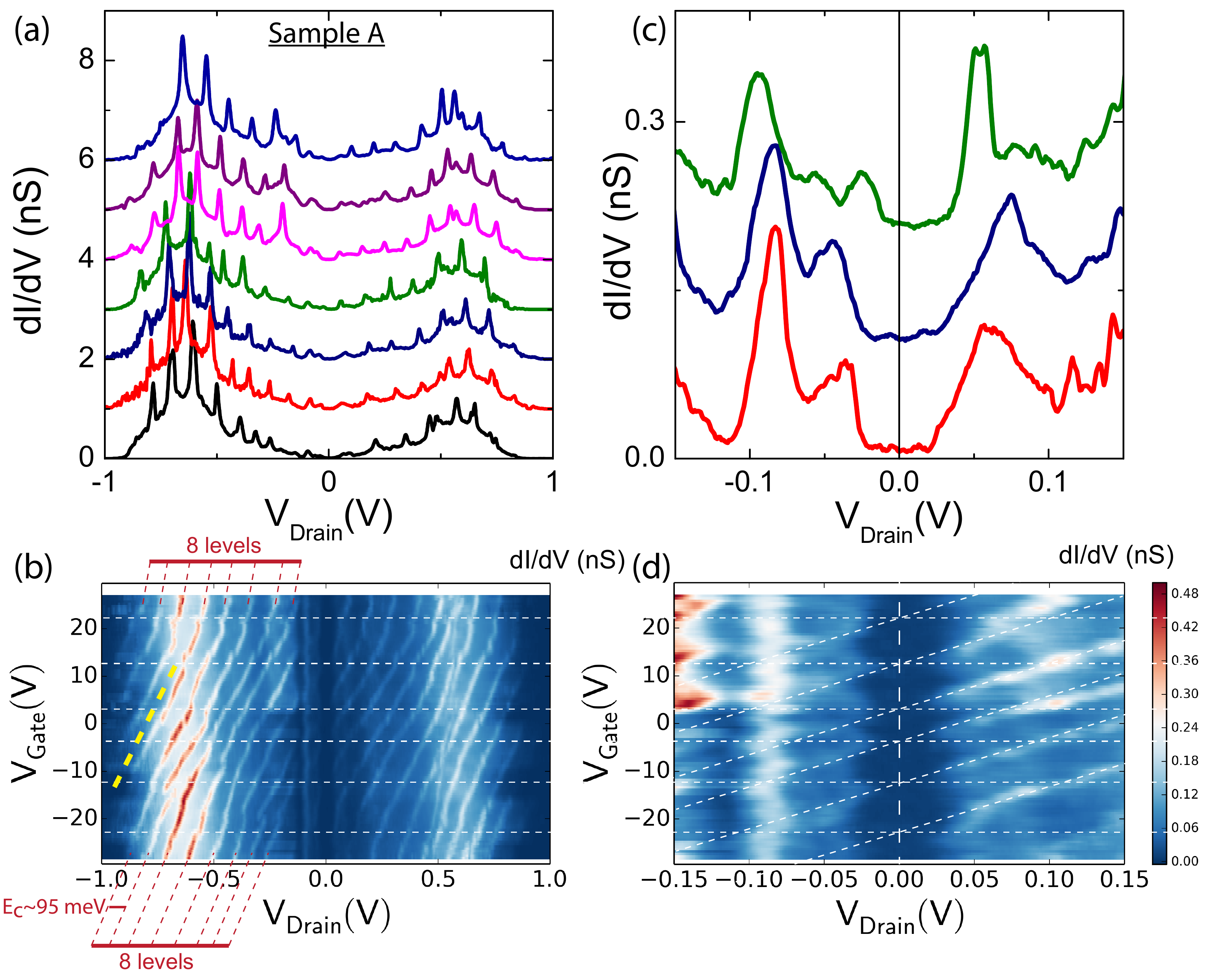}
		\caption{\label{Fig3} $dI/dV$ curves for sample A, plotted on panel (a) for $V_{Gate}=-27,-17,-7,0,7,17,27$~V  (bottom to top) and shown on the color plot (b) as function of drain and gate voltage, measured at $T=5$~K. The red dashed lines highlight the eight Coulomb peaks of the 1S$_h$ excited level. The yellow dashed line is used to calculate the back-gate lever arm $\alpha_C$. Zoom on the $dI/dV$ curves at low drain bias, from $V_{Drain}=-0.15$~V to $V_{Drain}=+0.15$~V, plotted on panel (c) for $V_{Gate}=-17,-7,0$~V and shown on the color plot (d). The white horizontal dashed lines highlight the gate voltage where the number of electrons in the QD is changed by one. This zoom shows that the gap at low bias cannot be lifted by the gate voltage. On panels a) and c), the curves have been shifted up for clarity.
		}
	\end{center}
\end{figure}

Because PbS has the rock-salt crystal structure and, as a result, has direct band gaps at four equivalent L points in the Brillouin zone\cite{Kang1997}, the excited levels 1S$_e$ and 1S$_h$ are 8 times degenerated, after taking into account the spin degeneracy. In the shell-filling regime, this implies that up to 8 peaks separated by the Coulomb energy should be observed in the conductance curves. Fig.~3a shows the $dI/dV$ curves for sample A as function of gate voltage, shown on the color plot Fig.~3b. At any gate voltage, exactly 8 conductance peaks can be clearly distinguished as function of drain voltage. This implies that the injected electrons are indeed populating the 1S$_e$ and 1S$_h$ levels of the QD. The fact that excitations occur primarily in one direction is due to asymmetric tunnel barriers\cite{Kouwenhoven2001}. For this reason, we can assume that the applied voltage difference $V_{app}$  across the electrode-dot-electrode system is mostly dropping on a single junction, i.e. the voltage division $\eta=V_{Drain}/V_{app}\sim1$\cite{Banin2003}, which implies that the observed gap is close to the real QD gap. Fig.~3b shows that the Coulomb peaks are shifted with the gate bias and eventually cross zero-energy, where the number of electrons in the QD changes by one, and leads to the apparition of Coulomb diamonds, as shown on the zoom at low bias, Fig.~3d. Such behavior was also observed for sample B, shown Fig.~4. For this sample, the Coulomb energy $E_c\sim 50$~meV and so the QD diameter is $2\times r \sim 16$~nm. Because of this larger diameter, excitations levels are broad and not clearly apparent for this sample. However, as seen below, this sample allows observing clear phonon sub-bands.

Before turning to this, a few remarks are in order. The calculated capacitance between a sphere of radius $r$ and a metallic plane at the gate distance $d=300$~nm gives $C_{sp}/e=5.3$~V$^{-1}$ for sample A and $C_{sp}/e=10.2$~V$^{-1}$ for sample B\cite{Supp}. We find for the experimental values $C/e= 0.1$~V$^{-1}$ for sample A and $C/e= 2.5$~V$^{-1}$ for sample B. These values are smaller than the theoretical value because of the screening effects due to the electrodes, which depend on the exact position of the QD with respect to the electrodes. One can see, for sample A, that the back-gate lever arm is different for the Coulomb and the excited levels (1S$_e$,1S$_h$). While the lever arm for the Coulomb peak is $\alpha_C = \delta E_c/\delta V_{Gate} \sim 0.0085$, the excitation peaks are barely shifting with the gate. This can be understood as a consequence of the good screening properties of PbS which has a large static dielectric coefficient. This effect is not important for the present discussion on the electron-phonon coupling. Finally, the observation of Coulomb diamonds is usually expected in metallic nanoparticules or in semiconducting QDs where the Fermi level has been driven in the conductance or valence band with the gate voltage. Even if the applied gate voltage is not sufficient to push the excited levels across zero bias, the broadening of excited levels is sufficient to produce a residual density of states within the semiconducting gap, allowing the QD to effectively behave as a metallic nanoparticle. This is consistent with the recent STM observation of midgap states in PbS QDs\cite{Diaconescu2013} and transport measurements in PbS QDs thin films\cite{Nagpal2011}.


\begin{figure}[ht!]
	\begin{center}
		\includegraphics[width=8cm]{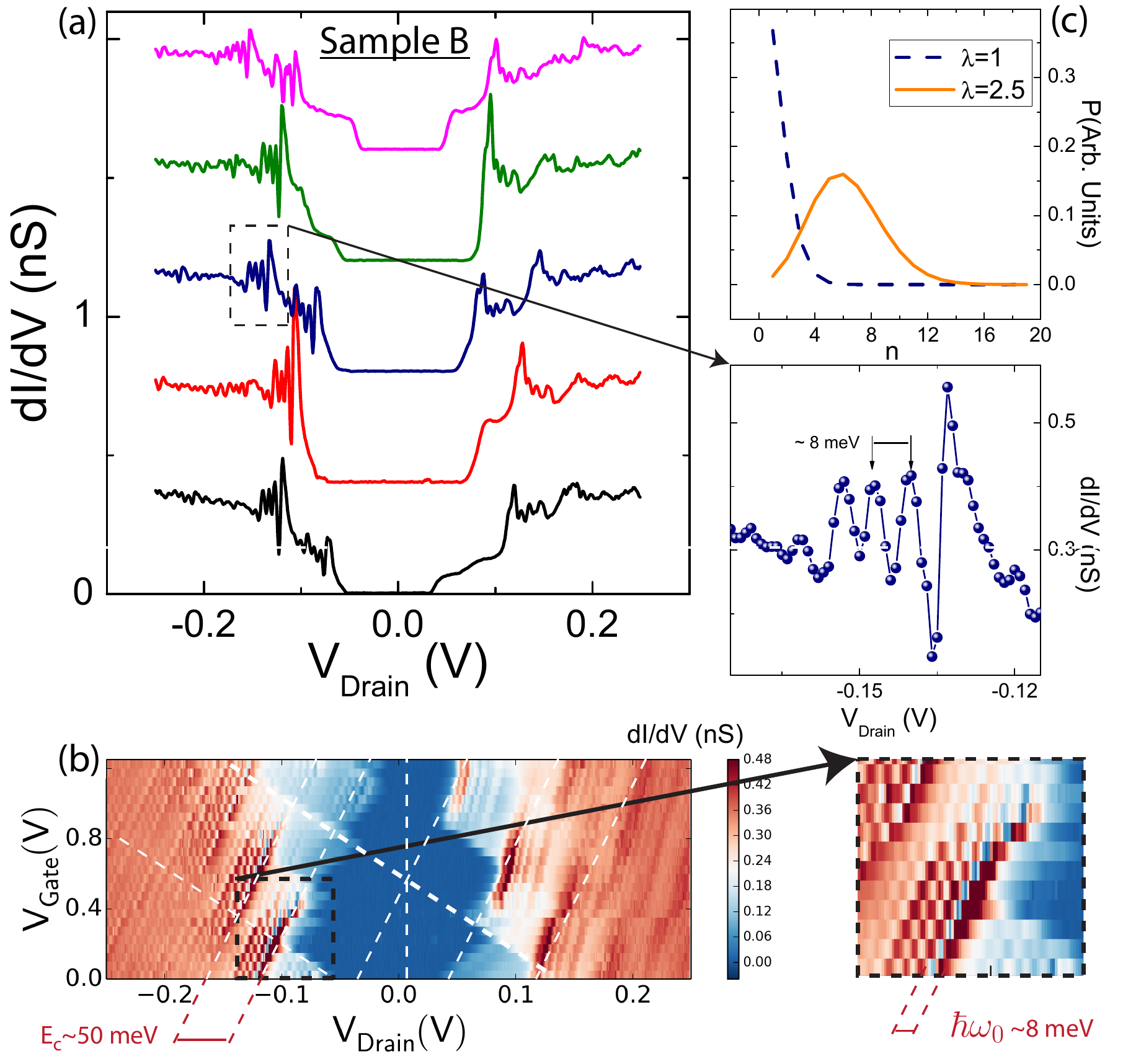}
		\caption{\label{Fig4} $dI/dV$ curves for sample B, plotted on panel (a) from $V_{Gate}=0$~V (bottom)to $V_{Gate}=1.2$~V (top) and shown on the color plot (b) as function of drain and gate voltage, measured at $T=5$~K. These panels show that the gap at low bias cannot be lifted by the gate bias. The zoom on the $dI/dV$ curve at $V_{Gate}=0.4~V$ and the zoom on the color plot show that a single Coulomb peak is formed of phonon sub-bands separated by the phonon energy $\hbar \omega_0 \sim 8$~meV. (c) Theoretical amplitude, Eq.~\ref{Eq:FC}, of the FC peaks as function of the number of emitted phonons for two values of the Huang-Rhys factor $\lambda=1$ (dashed line) and $\lambda=2.5$ (continuous line). At large $\lambda$, the matrix element goes to zero for small n, indicating the FC blockade.
		}
	\end{center}
\end{figure}

As we have seen, the degeneracy lifting effect of the Coulomb energy is the main origin for the broad peak observed Fig.~2. However, the inset of Fig.~2 shows that a single Coulomb peak has a width $\sim20$~meV which is still much broader than the thermal smearing at $T=5$~K. Similar broadening were observed in STM spectra on CdSe\cite{Sun2009} and PbS\cite{Diaconescu2013}.

A zoom at the Coulomb peaks measured on sample B, Fig.~4, clearly shows that the Coulomb peak is constituted of sub-bands separated by an energy of $\sim 8$~meV. These peaks can also be observed for sample A, but with lower resolution. These peaks are equally spaced and strongly resemble the expected response when the electron level is coupled to phonon modes\cite{Wingreen1989,Mitra2004,Delerue2004}. This behavior has been observed previously in STM spectroscopy of CdSe QDs\cite{Sun2009}, in molecules\cite{Park2000,If2002,Qiu2004} and nanotubes based QDs\cite{Sapmaz2006,Leturcq2009}.

The coupling of electronic levels with vibrational modes can be described in terms of the FC model\cite{Wingreen1989,Mitra2004,Delerue2004}. In the case of a single phonon mode $\hbar \omega_0$, the FC theory gives for the transition probability :

\begin{equation}\label{Eq:FC}
X_{0n}^2=|<0|X|n>|^2=\frac{e^{-\lambda^2}\lambda^{2n}}{n!}
\end{equation}

between a state with 0 phonons and a state with $n$ phonons where $\lambda$ is the electron-phonon coupling strength, also called the Huang-Rhys factor.

In bulk PbS, the energy of the zero-wave-vector
($\Gamma$-point) transverse-optical phonon is 8.1 meV as observed through far-infrared absorption\cite{Krauss1996} spectroscopy and Raman spectroscopy\cite{Krauss1997,Krauss1997a}. Furthermore, vibronic quantum beats have also been observed in femtosecond optical spectroscopy\cite{Krauss1997a,Bylsma2012} of PbS QDs. 

\begin{figure}[ht!]
	\begin{center}
		\includegraphics[width=8cm]{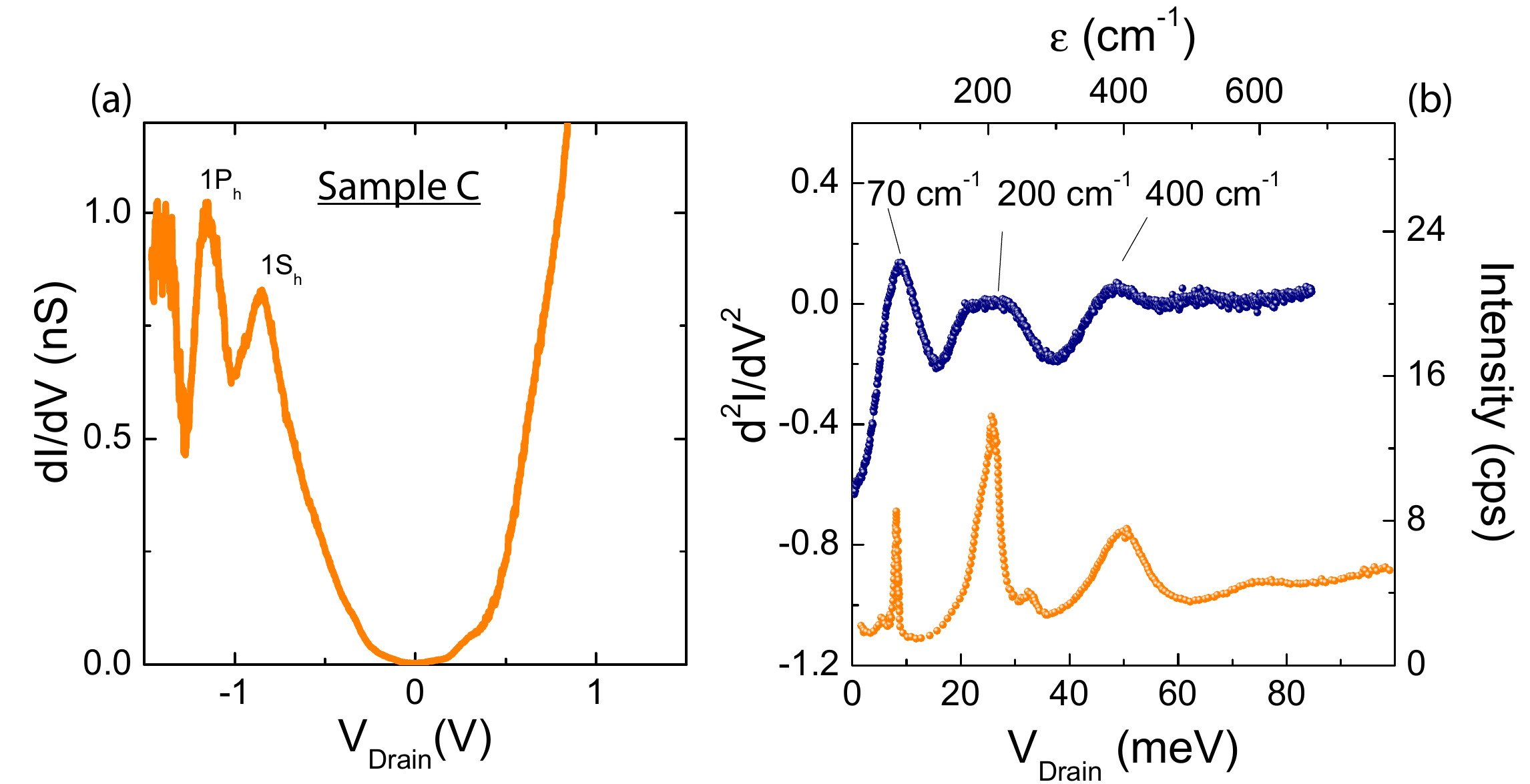}
		\caption{\label{Fig5} (a) $dI/dV$ curves for sample C showing the excited hole levels,  1S$_h$,  1P$_h$, measured at $T=5~K$. Note the absence of the Coulomb peaks in this shell-tunneling regime. (b) Inelastic ETS $d^2I/dV^2$ showing three lowest phonons mode compared to the Raman spectrum extracted from Ref.\cite{Krauss1997}.
		}
	\end{center}
\end{figure}

Phonon modes can also be observed in the inelastic ETS\cite{Hansma1982}. These low energy modes could not be observed in samples A and B because of the Coulomb gap at low bias. However, one of the studied sample was in the regime of shell-tunneling and, consequently, did not present Coulomb blockade effects, Fig.~5a. The absence of the sharp Coulomb blockade peaks does not allow the observation of the phonon sub-bands, however, the absence of the gap at zero bias allows measurements of the inelastic ETS $d^2I/dV^2$, shown Fig.~5b. This last spectrum shows the first three optical phonon modes at the position expected from Raman spectroscopy\cite{Krauss1997}.



Returning to samples A and B, one observes, Fig.~3ab and Fig.~4ab, respectively, that a gap remains at low bias at any gate voltage. Given the signature of strong electron-phonon coupling observed in these PbS QDs, a FC blockade could be at the origin of this low bias suppression of conductance\cite{Koch2005,Koch2006}. While the Coulomb blockade can always be lifted at appropriate gate voltage values, the FC blockade cannot be lifted by a gate bias, which is a distinguishing feature of the FC blockade. The observation of FC blockade in a tunneling experiment has been observed previously in GaAs based QDs\cite{Weig2004}  and carbon nanotubes based QDs\cite{Sapmaz2006,Leturcq2009}. The FC blockade originates from the behavior of the FC matrix element $X_{0n}$. When tunneling on the QD, the electron shifts the equilibrium coordinate of the QD by an amount proportional to the Huang-Rhys factor $\lambda$.  As the overlap between states of different phonons occupation is exponentially sensitive to this geometrical displacement, the ground-state to ground-state transition is exponentially suppressed for strong electron-phonon coupling. 

 
For equilibrated phonons, this suppression dominates until the bias voltage is high enough, $eV\sim \lambda^2 \hbar \omega_0$\cite{Koch2005,Koch2006}, to escape from the blockade regime by transitions from zero phonons to highly excited phonon states.
From the observed gap values for sample A ($\sim 25$~meV) and sample B  ($\sim 50$~meV), we find that the electron-phonon coupling constant is in the range $\lambda \sim 1.7-2.5$, which is very large, of the order of the Huang-Rhis factor obtained from Raman scattering experiments\cite{Krauss1997}. While there is no consensus on the effects of quantum confinement on electron-phonon coupling, see. Ref. \cite{Sagar2008} for a review, it has been suggested that a large electron-phonon coupling in QDs could be the consequence of trapped charges at the surface of QDs\cite{Krauss1997} or polaronic effects that would arise as a consequence of the discrete electronic levels\cite{Ferreira2003}.

To summarize, we found that the elastic and inelastic ETS of PbS QDs is characterized by signatures of strong electron-phonon coupling. In the shell-tunneling regime, three phonon modes can be observed in the \emph{inelastic} ETS $d^2I/dV^2$. In the shell-filling regime, where the Coulomb blockade peaks are observed, the lowest energy phonon mode leads to the apparition of sub-bands that can be observed in the \emph{elastic} ETS $dI/dV$.
In this regime, we observe that the Coulomb blockade cannot be lifted at any gate voltage, which is likely the consequence of FC blockade. Thus, this first report of the observation of FC blockade induced by coupling of electrons to optical phonons teaches us that using QDs with low electron-phonon coupling should help improve electronic transport in QDs thin films.

We thank M. Rosticher for his technical support with the clean room work. We acknowledge support from ANR grant "QUANTICON" 10-0409-01, ANR grant "CAMELEON" 09-BLAN-0388-01,  Region Ile-de-France in the framework of DIM Nano-K and China Scholarship Council.

\bibliography{Bibliography}

\end{document}